\documentclass[12pt]{article}
\pdfoutput=1
\usepackage{amsmath}
\usepackage{amsfonts}
\usepackage{amsthm}
\usepackage{amssymb}

\usepackage[pdftex]{graphicx,color}
\usepackage[pdftex]{hyperref}
\newcommand{\sech}{\operatorname{sech}}

\numberwithin{equation}{section}

\begin{document} 

\title{Phase space methods and psychoacoustic models in lossy transform coding}
\author{Matthew C.~Cargo\\
1615 Prince St. Apt. 9, Berkeley, CA, 94703\\
(510)704-8096, \sf{mcargo@gmail.com}}

\maketitle
\begin{abstract}
I present a method for lossy transform coding of digital audio that uses the Weyl symbol calculus for
constructing the encoding and decoding transformation.  The method establishes a direct
connection between a time-frequency representation of the signal dependent threshold of
masked noise and the encode/decode pair. The formalism also offers a time-frequency measure
of perceptual entropy.
\end{abstract}

\maketitle
\bibliographystyle{IEEEtran}
\pagestyle{plain}

\section{Introduction}

In lossy transform coding, a signal is transformed before re-quantization, and
then partially recovered by applying the inverse transformation. In
perceptual codecs, the goal is to make the necessarily introduced noise
imperceptible. Mathematically, let $\psi$ be the original signal, $\hat K$ (for 
``key'') be a linear
transformation, and $\hat L =\hat K^{-1}$ (for ``lock'') be its inverse. Then
\begin{equation}
\psi'=\hat K \hat Q\hat L\psi
\end{equation}
is the recovered signal.  Here, $\hat Q$ is a quantization operator.
It is not a linear operator,\footnote{Blocking artifacts are difficult
to analyze precisely because $\hat Q$ is not linear.}
but we can often model $\hat Q$ as
introducing noise:
\begin{equation}
\hat Q\phi=\phi+a_Q X,
\end{equation}
where $X$ is a time series of uniformly distributed independent random variables
on the interval $(-1/2,1/2)$, and the constant $a_Q$ is determined by the 
quantization scale.  Hence, for the reconstituted signal as above, 
the introduced noise is
\begin{equation}
a_Q \hat K X.
\end{equation}
The noise is no longer white, but rather shaped by the operator
$\hat K$.

A good psychoacoustic model will determine whether this noise is masked by $\psi$, and a good lossy
encoding algorithm will choose $\hat K$ so that it minimizes the combined storage requirements of
the key $\hat K$ and the encoded signal $\hat Q\hat K^{-1}\psi$,
subject to the constraint that the introduced noise cannot be heard. 

In this paper, I extend the types of transformations to include pseudo-differential operators. In 
language of signal processing, a pseudo-differential operator on a 
sampled signal is a matrix with limited extent off its diagonal and
a limited rate of change along the diagonal; one could
also call it a slowly evolving filter.  As far as I can tell,
the community has not used pseudo-differential operators in
transform codecs, because, I would guess,
they are not diagonal in any standard basis and are therefore more
difficult to invert.\footnote{For example, there is the following statement in
\cite{herre}: ``In order to perform well for most signals,
 however, the processing has to be applied to different parts of the frequency 
spectrum independently, since transient events are often present only in certain 
portions of the spectrum. This can be done using more complex hybrid filterbanks 
that allow for separate gain processing of different spectral components. In general, 
however, the interdependencies between the gain modification and the coder's perceptual 
model are often difficult to resolve.''}  The phase space theory of these operators, below,
resolves the presumed difficulties and brings pseudodifferential operators into the
realm of practical transforms.\footnote{Be advised that the method is the subject
of a provisional patent application to the United States Patent Office.}

\section{The Weyl symbol}
A symbol correspondence is a bijection between operators (here,
on signals) and functions on the corresponding classical phase space 
(here, the time-frequency plane). The canonical symbol correspondence
is the Weyl \cite{Weyl} symbol. It enjoys many properties
that entitle it to be called ``the'' phase space representation of
an operator, and is defined as follows. If $\hat A$ is an operator
with $t$-space matrix elements $\langle t_1\vert \hat A\vert t_2\rangle$,
its Weyl symbol $(s\hat A)$ is a function of $t$ and $f$ defined by
\begin{equation}
(s\hat A)(t,f)=\int ds\, e^{2\pi i f s}\langle t+s/2\vert \hat A
\vert t-s/2\rangle.
\end{equation}
That is, the Weyl symbol is the Fourier transform of the matrix
in its difference variable. 

Some examples and properties of $s$:
\begin{enumerate}
\item If $\hat I$ is the identity operator, then $(s\hat I)(t,f)=1$.
\item If $\hat A$ is diagonal in the $t$-representation with 
$\langle t\vert\hat A\vert t\rangle=a(t)$, then $(s\hat A)(t,f)=a(t)$.
If $\hat A$ is diagonal in the $f$-representation, with diagonal
elements $b(f)$, then $(s\hat A)(t,f)=b(f)$.  This parallel is part
of a larger metaplectic covariance of the formalism.
\item If $\psi$ is a signal, then we can form the one dimensional
projection operator $\hat A=\vert \psi\rangle\langle\psi\vert$.
The Weyl symbol of this operator is called the Wigner \cite{Wigner} function, 
which, in signal processing, is a type of spectrogram.  It is
typically a rapidly varying function on phase space.
Generally speaking, the Wigner function contains too much information
to be of use for our purposes. However, various smoothings of it are valuable.
\end{enumerate}

The Weyl symbol allows us to regard operators as functions on phase space.
However, in order to use these functions, we need to know what happens to operator multiplication.
It becomes the {\it star product,} defined by
\begin{equation}
(s\hat A)\star (s\hat B)=s(\hat A\hat B).
\end{equation}
The star product acts generally through an integral kernel
$\Delta$,
called the trikernel:
\begin{equation}
\begin{split}
((s\hat A)\star(s\hat B))(z)&=
\int dz_1\, dz_2\, (s\hat A)(z_1)(s\hat B)(z_2)\\
&\times \Delta(z,z_1,z_2),
\end{split}
\end{equation}
where $z=(t,f)$ etc, and $\Delta$, not needed elsewhere, is an exponential
of the area spanned by the triangle with vertices $(z,z_1,z_2)$. 
The trikernal formula is useful in some contexts, such as when $\hat B$ is
a Wigner function, but is often too complicated to be of value. However, 
there is a class of functions we will call
slowly varying, for which a simpler expression holds. (These functions
are pseudo-differential operators.)

If $A$ and $B$ are slowly varying symbols, then their star product can be 
expanded as series of bidifferential operators, called the Moyal \cite{Moyal} star
product:
\begin{equation}
(A\star B)(t,f)=A(t,f)\exp\left(\frac{i}{2}\stackrel{\leftrightarrow}{J}\right)
B(t,f)
\end{equation}
where the ``Janus'' operator 
\begin{equation}
\stackrel{\leftrightarrow}{J}=
\frac{1}{2\pi}\left(
\frac{\stackrel{\leftarrow}\partial}{\partial t}
\frac{\stackrel{\rightarrow}\partial}{\partial f}
-
\frac{\stackrel{\leftarrow}\partial}{\partial f}
\frac{\stackrel{\rightarrow}\partial}{\partial t}
\right).
\end{equation}
Here, derivatives topped with left (right) arrows act to the left (right, resp.)
Conversely, if the Moyal series converges for
$A\star B$, then they are slowly varying. 

In the case of extremely slowly varying functions, the star product is
well-approximated by its leading term, the ordinary product.  This
is important: it means for the right operators, the Weyl
transform maps complicated operator multiplication 
to simple ordinary multiplication.

How do we know {\it a priori} whether a function is slowly varying?  One
way is to consider sets of functions having rigorous bounds
on the ratio of higher terms in the Moyal series to the leading term.
On such set is the set of bounded variation.
We say $A(t,f)$ is of bounded variation, with length scales $(a_t,a_f)$, if
\begin{equation}
\left|
\frac{1}{A}
\frac{\partial ^n\partial^m A}{\partial t^n\partial f^m}
\right|\le \frac{1}{a_t^n}\frac{1}{a_f^m}.
\end{equation}
With this definition, it is easy to prove that if $A$ and $B$ are
of bounded variation and $2\pi a_t a_f\gg 1$,
then the Moyal series converges. In other words, the area of
the characteristic scale of variation must be much larger than a Planck cell.
This is a direct consequence of time-frequency uncertainty.

We can create functions of bounded variation by using a $\sech$ kernel: if 
\begin{equation}
\label{sech_kernel}
k(t,f)=\frac{1}{4 a_t a_f}\sech\left(\frac{\pi}{2} \frac{t}{a_t}\right)
\sech\left(\frac{\pi}{2} \frac{t}{a_f}\right),
\end{equation}
and $A(t,f)$ is a positive-valued function, then
the convolution $k\circ A$ is of bounded variation, with
scales $(a_t,a_f)$.\footnote{The normalization is
chosen so that the convolution tends to the identity operator
as $(a_t,a_f)\to 0$.}  Note the convolution is on the whole phase space,
rather than just in $t$ or $f$.

Finally, we introduce an important formula for the symbol of
a function of an operator (the sofoo formula), i.e., $s f(\hat A)$.  
For example, given $\hat A$ or its symbol $A=s\hat A$, we will need to 
know $s \hat A^{1/2}$ and
$s \hat A^{-1}$. Fortunately, the subject was considered at
length by Gracia-Saz \cite{graciasaz}. The general formula is quite complicated and
expressed in terms of a series of diagrams.  For our purposes, the
important facts are these: first, which follows directly from the
Moyal product, that
\begin{equation}
s f(\hat A)=f(s\hat A)+\operatorname{h.o.t.}, 
\label{sfa}
\end{equation}
and the higher order terms at right involve successively more derivatives of 
$s\hat A$; second,
that those derivative terms contain $t$ and $f$ derivatives
in pairs. Thus, if $A$ is of bounded variation, then
the series for $sf(\hat A)$ is well-approximated by its first term.
However, even $A$ is not of bounded variation, then the series
might still converge; in particular, we might imagine varying
over phase space the smoothing scales $(a_t,a_f)$ while maintaining a constant
product.

The first correction term to Eq.~(\ref{sfa}) is
\begin{equation}
\begin{split}
-\frac{1}{4}\frac{1}{(2\pi)^2}\left(\frac{1}{2}\frac{f''(A)}{2!}(A_{tt}A_{ff}
-A_{tf}^2)\right.\\\left.+\frac{f'''(A)}{3!}(A_t^2A_{ff}+A_f^2 A_{tt}-2A_tA_f A_{tf})\right).
\end{split}
\end{equation}
The diagrams in \cite{graciasaz} and \cite{cargo_thesis} help express 
this result more economically. This formula may find application below
when the first term alone is not accurate enough.\footnote{The situation 
is worse for other symbol correspondences, e.g., the normal ordered
symbol for which $s^{-1}(tf) =-i /(2\pi) t\partial_t$. For those symbols,
the first correction term in the sofoo formula is of lower order; in terms
of the scale of variation, it scales as $1/(a_t a_f)$ and is less easily ignored.}

\section{\label{section_noise}Phase space description of noise.}
Fully understanding a times series of random variables $Y(t)$, 
requires knowing its entire joint probability distribution.
However, for most purposes, it suffices to study only the
two point correlation function, that is, the expectation of
a product at two times:
\begin{equation}
E(Y(t_1) Y(t_2)^*),
\end{equation}
assuming the variables have individual means of zero. This
expectation allow us to define a Hermitian noise operator $\hat N_Y$, whose
$t$ space matrix elements are as above, i.e., 
$\langle t_1\vert \hat N_Y\vert t_2 \rangle=E(Y(t_1)Y(t_2)^*)$.

We will characterize a given noise operator by its Weyl symbol.
For example, the noise operator for white noise $X(t)$ with unit variance 
is the identity operator, and hence its symbol is unity---which makes
sense.  The random variable series for time localized noise
is $Y=WX$, where $W$ is a window.  It follows that the
Weyl symbol for this noise is just $W(t)^2$.  Colored noise,
on the other hand, is defined by its Fourier transform:
$\mathcal{F}Y=W X$, and its Weyl symbol is $W(f)^2$.

The converse problem is to {\it produce} noise with a given (Hermitian) noise
operator $\hat N$, and the solution is simple: defining
$Y=\hat N^{1/2}X$ gives the desired noise, since
\begin{equation}
E(\vert Y\rangle\langle Y\vert)
=\hat N^{1/2} E(\vert X\rangle\langle X\vert)\hat N^{1/2}
=\hat N.
\end{equation}
Here we use for white noise that $E(\vert X\rangle\langle X\vert)=\hat I$. 
As a practical matter, finding the square root of a given
operator may not be so easy. However, if the operator has
a slowly varying Weyl symbol $N$, the matter is straightforward: 
$\hat N^{1/2}=s^{-1}s\hat N^{1/2}\approx s^{-1}N^{1/2}$; that is,
we simply take the square root of the noise operator's symbol, and convert
it back to an operator using the inverse symbol. Note
that this procedure does require that $N$ is a positive function.

As a psychoacoustical matter, can we always describe noise by
a slowly varying operator? Consider, for example, noise 
created from white noise by applying rapidly varying operators. In 
particular, the one-dimensional projector 
$\vert\psi\rangle\langle\psi\vert$ onto
a signal $\psi$ is definitely not a slowly varying operator, and,
when applied to white noise, gives a signal proportional to
$\psi$ itself---the only randomness left is in the overall norm of
the signal.
Roughly speaking, the more rapidly an operator $\hat A$ varies, the
more structure it imparts to a white noise signal $X$, and the less
noisy the resulting signal sounds.

\section{Phase space setting for masking of noise.}

Psychoacoustical experiments of signals masking noise
are consistent with the hypothesis that the maximal
noise masked by a given signal is a
slowly varying noise operator. Masking experiments, except
for those done informally in the testing of compression
algorithms, are typically done in time or frequency, but not both.
The classic paper by Ehmer \cite{ehmer} shows masking curves of
noise by pure tones. The curves typically
peak at the tone frequency and fall off at a scale proportional
to the frequency itself, but faster toward decreasing frequency.
Temporal masking experiments show pre-masking rising to a certain
threshold under the signal, and decaying afterward.

We can generalize these results
to a broader hypothesis: For a given signal $\psi$, there
exists a noise operator $\hat M_\psi$, such that $\hat A X$ 
is fully masked by $\psi$ whenever $s(\hat A^2)$ is strictly
less than $M_\psi=s\hat M_\psi$.  In other words, $\psi$ generates a phase
space profile for the maximum allowed noise. 

This phase space profile must be related to the phase space profile of
$\psi$ itself---but how? We have already mentioned that the Wigner
function is typically not useful for phase space analysis.  To begin,
it is not slowly varying. Moreover, it sometimes falls below zero.
We can guess, however, that the masking noise profile of $\psi$ 
might be related to a smoothing of the Wigner function over phase space.
One particular smoothing of the Wigner function gives 
another well-known phase space distribution---the coherent state representation.

The coherent state representation $C_\psi(t_0,f_0)$ is defined as follows. 
Using the moving gaussian window $w_{t_0}$ with width $a$ defined by
\begin{equation}
w_{t_0}(t)=\exp\left(-\frac{(t-t_0)^2}{2 a^2}\right)
\end{equation}
we define
\begin{equation}
C_\psi(t,f)=\vert (\mathcal{F} w_t\psi)(f)\vert^2,
\end{equation}
where $\mathcal{F}$ is the Fourier transform. The coherent state
representation depends on the parameter $a$, making it less
canonical than the Wigner function; on the other hand, the finite
width of the window makes it much easier to calculate.  The coherent
state representation, regarded as a symbol of an operator, is
not slowly-varying, but it does vary more slowly than the Wigner function,
and it is never negative.

Our hypothesis, then, is that $s\hat M_\psi$ is related to a smoothing
of $C_\psi$ (which is itself a smoothing of $s(\vert\psi\rangle
\langle\psi\vert)$) with normalization and width parameters determined 
by listening tests. The future full theory will take into account 
different masking widths at
different frequencies as well as the statistical properties of $C_\psi$,
in order to account for the known assymmetry between the masking of
noise by tones and the masking of noise by noise.

In the meantime, I have explored simplified theories that, though yielding 
sub-maximal phase space noise 
thresholds $M_\psi$, nevertheless condemn to obscurity noise operators 
whose symbols fall below them. I will call such noise operators
noise-confining operators; the goal for more sophisticated
psycho-acoustical models will be an algorithm for generating the maximal noise-confining
operator---however, as we shall see, a sub-maximal noise-confining
operator can still be useful.

Finding a noise-confining operator is straightforward.
For a signal $\psi$, I smoothed $C_\psi$ by convolving it with
the $\sech$ kernels of Eq.~(\ref{sech_kernel}) in order to produce 
an easily manipulated
function $S_\psi$ of bounded variation. I used width parameters
suggested by masking experiments. To test the theory, I took 
$s^{-1}(S_\psi^{1/2})$, and, as explained in section \ref{section_noise},
applied this operator to a noisy signal $x$ (a realization
of the uniformly distributed random variable series $X$). 
I then listened to
\begin{equation}
\label{psymodel}
\psi+\alpha\, s^{-1}(S_\psi^{1/2})x
\end{equation}
and increased $\alpha$ to the threshold at which the above began to sound
different from $\psi$.  By repeating this for different signals,
and choosing the smallest $\alpha$, I became confident that 
\begin{equation}
\hat M_\psi = \alpha^2\, s^{-1}(S_\psi^{1/2})^2,
\end{equation}
did indeed describe a noise-confining operator.

\section{Phase space codec}
In the previous section, I introduced an explictly phase phase space 
setting for the signal dependent threshold of noise,
and we will now use it to design a lossy transform codec. 
First, a note about normalization. I will assume
that the original signal $\psi$ and the encoded signal
$\psi_{encoded}$ are both quantized on a unit scale.
\footnote{Since $\psi_{encoded}$ is determined from the psychoacoustical
model, this method is inherently variable bit rate.}
The quantization noise $X$ present in $\psi_{encoded}$ may,
under certain conditions, be described as uniformly distributed
on the interval $(-1/2,1/2)$ with variance $1/12$.
We have, for the encoded and restored signals, that
\begin{equation}
\label{codec}
\begin{split}
\psi_{encoded}&=\hat Q \hat K^{-1}\psi\\
\psi_{restored}&=\hat K\psi_{encoded}\approx\psi+\hat K X.
\end{split}
\end{equation}
Comparing Eqs.~(\ref{codec}) and (\ref{psymodel}), we set
\begin{equation}
\hat K =s^{-1}(M_\psi^{1/2})
\end{equation}
so that the noise introduced into $\psi_{restored}$ is just at the threshold
measured by the psychoacoustical model. 

I now present the argument showing how $\hat K$ reduces the average bit content of 
$\psi_{encoded}$. I use an empirical
observation that the values taken by $\psi_{encoded}$ are uniformly
distributed over its range, but the argument does generalize easily to 
more general distributions. If this assumption is true, we can
estimate the average size of $\psi_{encoded}$ from its variance:
\begin{equation}
\label{mean_calc}
\begin{split}
\frac{1}{N}
\langle\psi_{encoded}\vert\psi_{encoded}\rangle&=
\frac{1}{N}
\operatorname{tr}(\vert \psi_{encoded}\rangle\langle
\psi_{encoded}\vert)\\
&=
\frac{1}{N}
\operatorname{tr}(\hat M_\psi^{-1/2}\vert\psi\rangle
\langle \psi\vert \hat M_\psi^{-1/2})\\
&=
\frac{\int dt\,df\,M_\psi^{-1}s(\vert \psi\rangle\langle\psi\vert)}
{\int dt\,df}
\end{split}
\end{equation}
Here $\operatorname{tr}$ takes the trace of its operand, and,
in the last line we have used the traciality property of
the Weyl symbol, namely, that
\begin{equation}
\label{traciality}
\operatorname{tr}(\hat A\hat B)
\propto \int dt\,df\, s\hat A\, s\hat B.
\end{equation}
On the right, it is important to note the ordinary product appears
rather than the star product.  The traciality property converts
the mean over $t$ space into a mean over phase space.
In the last of Eq.~(\ref{mean_calc}), we have divided by
the phase space volume as a formal way to avoid worrying about the
normalization factor in Eq.~(\ref{traciality}).
Now, in the numerator integrand, the slowly varying function
$M_\psi^{-1}$ appears next to the rapidly varying Wigner
function $s(\vert\psi\rangle\langle\psi\vert)$. To a good
approximation, then, we may replace the Wigner function by
its average value within the variation scale of $M_\psi^{-1}$.
This average is, of course, $S_\psi$. Thus, if we
are working with the simplified model where $M_\psi = \alpha^2 S_\phi$,
we find the expectation
\begin{equation}
E(\psi_{encoded}^2) = \alpha^{-2}\approx 100.
\end{equation}

Using the information theoretic definition of entropy we can convert this
into a bit rate. Since we have not yet used that $\psi_{encoded}$
is uniformly distributed, we can afford to make a more general
argument in which $\psi_{encoded}$, before quantization,
takes its values from a probability density $p(\psi)d\psi$.
Quantization casts its values into bins $i$ of width $q(=1)$, and the 
probability that $\phi$ falls within the $i$'th bin is
$P_i $, where
\begin{equation}
P_i=\int_{q(i-1/2)}^{q(i+1/2)} d\phi\, p(\phi)\approx p(i),
\end{equation}
where we have used $q=1$.
The entropy per sample is
\begin{equation}
\begin{split}
S&=-\sum_i P_i \log_2 P_i\\
&=- \sum p_i\log_2 p(i)
\approx -\int d\phi\, p(\phi)\log_2 p(\phi)
\end{split}
\end{equation}
Thus, if $p(\phi)$ is uniformly distributed with standard deviation $\sigma$
\begin{equation}
\label{uniform_sigma}
S=\log_2 \sigma+\log_2 2\sqrt{3}
\end{equation}
Or, if $p(\phi)$ is
gaussian, 
\begin{equation}
\label{gaussian_sigma}
S=\log_2\sigma+\log_2 \sqrt{e\pi}
\end{equation} 
Now, $\sigma$ itself is obtained from $\phi$, leading to
\begin{equation}
\label{bit_rate}
\sigma^2=A_t\left[M_\psi^{-1}s(\vert\psi\rangle\langle\psi\vert) \right],
\end{equation}
where $A_t$ denotes a phase space average over a time scale
equal large enough to quell rapid variations in the result.
This formula, when plugged into either Eq.~(\ref{uniform_sigma})
or Eq.~(\ref{gaussian_sigma}), as appropriate, gives an
expression for the time-dependent number of bits consumed by $\psi_{encoded}$.
This formula evidently defines a phase space measure of perceptual entropy.

When, as in our simple model, $\sigma\approx 1/\alpha$, we find
\begin{equation}
S= \log_2 10 + \log_2 2\sqrt{3}\approx 5,
\end{equation}
so that the lossy stage of this encoding scheme takes no more than 5 bits per sample.

As for the coding, we may again employ the considerable
power of the sofoo formula and approximate
\begin{equation}
\hat L = \hat K^{-1}=s^{-1}(M_\psi^{-1/2}).
\end{equation}
That is, we simply invert the masking threshold $M_\psi$,
take its square root, and apply the inverse Weyl symbol.
This procedure ignores higher order terms in the
exact expression for $\hat K$'s inverse.
If this is not accurate enough, we can always write the operator more accurately
by using the higher order terms in the sofoo formula. 
(And this is okay, since time is the luxury of 
the coder.) 

\section{Summary of the codec so far}

Through listening tests, we refine a phase space theory
for the signal dependent threshold of noise. The outcome
is a mapping from $\psi$ to a noise operator $\hat M_\psi$.
We define a key operator $\hat K= s^{-1}M_\psi^{1/2}$
and send it off to a bit packing (entropy coding) algorithm 
for further compression.
Using the symbol of a function of an operator formula,
we define the lock operator $\hat L= s^{-1}M_\psi^{-1/2}$,
apply it to $\psi$, quantize the result, and deliver it also for bit 
packing.  This is the coding.
As for decoding, we unpack the key and the encoded signal and then
apply the key operator to it. 

\section{Practical issues and modifications}
In this section, we introduce two modifications which would have cluttered the earlier presentation.

Existing perceptual codecs, in addition to exploiting masking phenomena,
also use that much of the high frequency content is irrelevant
because we cannot hear it anyway.
This fact is easy to put into the phase space framework.  Let $\hat{H}$ be the noise operator 
for the frequency dependent threshold of human hearing, i.e., the loudest
colored noise that cannot be heard in any circumstances. We can then
add $\hat{H}$ to $\hat M_\psi$ in Eq.(~\ref{psymodel}) without changing
how $\psi$ sounds.  This suggests 
we take $\hat K = s^{-1}\left(M^{1/2}_\psi + \hat H^{1/2}\right).$ 
However, examining the formula for the $\psi_{restored}$, we see
that this key introduces noise that, though inaudible, 
is independent of the signal itself, meaning that it carries no
information. I have found that it works well
to keep the $\hat H$ term in the lock, but drop it from the key.  Two
choices that work well for the lock are
\begin{equation}
\begin{split}
s\hat L &= \frac{1}{M_\psi^{1/2}+ H^{1/2}}\\
s\hat L &= \frac{M_\psi^{1/2}}{M_\psi+ H}
\end{split}
\end{equation}
If we use these locks, then even in the not-quantized case, the restored signal is different from
the original. In the second lock above, it becomes
\begin{equation}
\psi_{restored} = s^{-1}\left(\frac{M_\psi}{M_\psi + \hat H}\right)\psi .
\end{equation}
This expression bears similarity to a Wiener filter in
\cite{kirschauer}.

Using this type of lock can significantly increase the subsequent lossless compressibility
of $\psi_{encoded}$. However, the improvement is not the same for all signals; those
with significant high frequency content retain their original compressiblity.

This brings us to the second deviation from the prior setup. So far, we have
always written $\psi_{encoded}$ in the time domain, but this is a problem
because $\psi_{encoded}$ is not as compressible there.  Even lossless 
compressors designed for the time domain, such as ones using linear prediction coefficients, 
do not perform as well as LZ or Huffman encoding in the frequency domain.
I have therefore found it better to package the encoded signal
in DFT'd chunks. To avoid the errors caused by
quantizing twice, I delay the quantization of the encoded signal until
after it has been Fourier transformed.  This is valid because white noise
is white in both the time and frequency domain.  However,
one must be careful to use a suitably large FFT.  If the chunks are
too small, frequency localization in the DFTs can cause the quantization noise assumptions 
to break down and introduce a noticeable warble to the decoded signal.
I find that chunks of 512 or 1024 samples work best.  In this
format, standard compression programs (like gzip) reduce monophonic samples at
44kHz from 4 to 12 percent of their original size.  However, this
does not include the storing the key.

\section{Storing the key}

The key spectrogram in this method takes the place of scale factor side information
in standard lossy codecs. Na\"ive lossless compression of a sampled key\footnote{
Of course, {\it any} key is a sampled key in this method. I usually
sample the spectrogram at one half the variance of the coherent state window. I assume
readers in this field are familiar with the transition from the continuous case, which I have
presented here for its ease of elucidation, to the discrete case which occurs in practice.}
spectrogram yields
disappointing results. Even though it is a smooth object with variations on the order
of 40 times the minimum uncertainty scale $\Delta f\Delta t = 1$, there is simply too much overhead in storing values at every point, or 
differences between them---I have tried just about everything. In compression of monophonic samples, no such lossless method made 
the key file take less than 10 percent of the sample size. I have realized that, in order to make this method competitive,
we must regard as only a suggestion that the key noise operator should be equal to the measured masking operator.
Of course, if we make the key bigger than that operator, we will no longer be in the noise confined regime.
Conversely, smaller keys sacrifice some of the available entropy. However, I have found that the key
can be stored at a fractional accuracy of 10 percent without substantially introducing audible noise or
degrading the compressibility.  

This lattitude allows us to store the key as an interpolated object where
the value at each knot is specified with only one byte. Specifically, I have used
an adaptive grid by allowing for variable time steps and then, for each selected time, sampling the slice at
a time-specific frequency step size. The step sizes for the adaptive grid are chosen as
large as possible subject to the constraint that linear spline over it
differs fractionally from the original key by no more than 10 percent.
The step size information, together with the values at the spline knots, comprise a much smaller object:
they reduce the overhead to less than 1 percent. One might think, given my earlier
emphasis on using functions with bounded variation so that the sofoo formula applies, that the obvious 
discontinuities introduced by this method would cause the whole framework to fall apart. However,
as is often the case in semi-classical analysis, we get more than we deserve using the final
results of na\"ive formal calculations: the method seems to work fine even with only piecewise smooth keys.  
If, however, in the future, these are found to introduce
artifacts, more sophisticated curve fits, such as cubic splines, could be developed, without, I think, sacrificing compressibility.
An alternative would be to store an interpolated key with the understanding that it would
be smoothed in a standard way after it is reconstituted; the practicality of such an approach would depend
on the spare computational overhead in the decode routine.

\section{Conclusion}

It is clear that we perceive sound in a time frequency plane, simply because we hear pitch and rhythm.  Thus,
any psychoacoustic theory should achieve its most natural form in phase space.  If I am
correct that the maximal noise masked by a given signal is always characterized by a slowly varying 
(pseudo-differential) noise operator, then this codec can exploit 
any valid psychoacoustical model. This makes it an attractive framework for directly translating
advances in the phenomenology of masking into better lossy data compression. It also offers an interesting perspective
on perceptual entropy.

The main practical concern is the processing load of the main decoding loop. In early, fairly unoptimized code, the decode runs 
faster than real time by a factor of two. The decode loop is $O(N)$, but the coefficient is rather 
large---on the order of $500$. Whether this loop can be implemented in real time on a portable device is beyond
my expertise.

I have not presented any suggestions for how this method develops in the stereophonic case.  It presents many new
and interesting issues, including psychoacoustical modeling of binaural masking effects and matrix-valued spectrograms.
I leave these matters to a future paper.  In the meantime, I can report that my early attempts at stereophonic compression---
in which I seperately calculate left and right smoothed spectrograms, use them to transform the left and right channels, and
then send the transformed mid and side channels to lossless compression---are transparent (informally) at 6 to 13 percent 
overall compression ratios.  It also works to form a single key from the mid channel and use it to encode both the mid and side channels.

On the whole, I am encouraged by the performance at this early stage. The method is quite young, and it clearly has many refinements and 
tweaks ahead of it. Beyond that, the formalism emphasizes the value of phase space methods in the treatment of noise, masking phenomena, 
and the measurement of perceptual entropy.%
\appendix
\section{Entropy of phase space noise}

I am not sure how the following argument fits in with the earlier entropy result of Eq.~(\ref{bit_rate}), but it is
yet another interesting application for the sofoo formula.  The entropy $S$ of a series of random variables $Y_i$ is 
\begin{equation}
S=-\int \prod_i dY_i\, P(Y) \log_2 P(Y),
\end{equation}
where $P(Y)=P(Y_1,\ldots,Y_N)$ is the joint probability density.
If $Y=\hat M X$, and $X$ is uncorrelated white noise, then we can use
that $P(Y)$ transforms as a one form to conclude that
\begin{equation}
S=-\int \prod_i dX_i \log_2\left( \operatorname{det}\hat M^{-1}\right).
\end{equation}
The $\log_2$ term being constant, we can integrate out the p.d.f for $X$ and resume as
\begin{equation}
\begin{split}
S&=\log_2\left( \operatorname{det}\hat M\right)=\log_2 2^{\operatorname{tr}\log_2\hat M}\\
&=\int dt\,df\, s(\log_2\hat M)\approx\int dt\,df\, \log_2 s\hat M,
\end{split}
\end{equation}
where we have used the traciality property and, in the last, assumed
that $\hat M$ is slowly varying. In this context, that $\psi$ can
tolerate the addition of noise $s^{-1}\hat M_\psi^{1/2}X$ means that
it belongs to an ensemble of identical sounding signals, and
its information content goes down by the above result.  This
result differs from the previous in that the original $\psi$ does
not appear, and the $\log_2$ is inside the integrand.

\section{Encoding already noisy signals}
Suppose $\psi$ already sounds noisy. Then, if the statistics of $\psi_{encoded}$ are correct, then the noisy part 
$\hat K \psi_{encoded}$ may suffice for a realistic sounding reconstitution of $\psi$.
The argument goes as follows. As usual,
\begin{equation}
\begin{split}
\psi_{encoded} & = \hat L \psi \\
\psi_{restored} & = \hat K \psi_{encoded} = \hat K\hat L\psi + \hat K X.
\end{split}
\end{equation}
We set 
\begin{equation}
\psi = \psi_{restored} = \hat K\hat L\psi + \hat K X
\end{equation}
and use that $\psi$ sounds noisy to approximate
\begin{equation}
\psi = \hat \kappa X,
\end{equation}
leading to 
\begin{equation}
\hat \kappa = \hat K\hat L\hat \kappa + \hat K.
\end{equation}
Inspection of this equation leads us to guess that
\begin{equation}
\begin{split}
\hat K & = \alpha_k \hat\kappa\\
\hat L & = \alpha_l \hat\kappa^{-1},
\end{split}
\end{equation}
so that
\begin{equation}
1 = \alpha_k\alpha_l +\alpha_k.
\end{equation}
This one equation does not determine these two proportionalities. 
We fix this by requiring that $\hat K$ be as large as possible, so that $\psi_{encoded}$ be as
small as possible. This implies $\alpha_l = 0$. Thus, when a signal is already noisy,
we can take $\hat L = 0$ and $s \hat K^2 \approx S_\psi$. In this extreme case, the
entire signal information is contained in the key. Of course, with $\hat L=0$, 
$\psi_{encoded} = 0$, and, in order that the reconstituted signal sound at all, we need
to dither white noise into $\psi_{encoded}$. Real signals will contain a fraction of 
noise and purer tones, so this extreme case will rarely actually occur; nevertheless, the
argument shows that noise can help us increase the overall key scale, and hence the compression
ratio for $\psi_{encoded}$. The argument also shows us another case where the lock
is not the key's inverse.%

\bibliography{codec_ref}
\end{document}